\documentclass[12pt]{article}
\usepackage{amsmath,amsfonts,setspace,indentfirst,hyperref}
\usepackage[english]{babel}
\usepackage{slashed}

\topskip \textwidth 480pt

\def\4{AdS_4 \times \mathbb{C}P^3}
\def\5{AdS_5 \times \mathbb{S}^5}
\def\cp3{\mathbb{C}P^3}

\def\be{\begin{equation}}
\def\ee{\end{equation}}
\def\ben{\begin{equation*}}
\def\een{\end{equation*}}
\def\ba{\begin{array}}
\def\ea{\end{array}}
\def\bn{\begin{aligned}}
\def\en{\end{aligned}}
\def\bnn{\begin{eqnarray*}}
\def\enn{\end{eqnarray*}}
\def\bsub{\begin{subequations}}
\def\esub{\end{subequations}}

\def\p{{\partial}}

\def\a{{\alpha}}
\def\b{{\beta}}
\def\G{{\Gamma}}
\def\g{{\gamma}}
\def\d{{\delta}}
\def\e{{\epsilon}}
\def\ehat{{\hat\epsilon}}

\def\h{{\eta}}
\def\th{{\theta}}

\def\k{{\kappa}}
\def\l{{\lambda}}
\def\m{{\mu}}
\def\n{{\nu}}

\def\r{{\rho}}
\def\s{{\sigma}}

\def\f{{\phi}}
\def\c{{\chi}}
\def\ps{{\psi}}
\def\w{{\omega}}
\def\W{{\Omega}}

\def\zb{{\bar{z}}}
\def\wb{{\bar{w}}}

\numberwithin{equation}{section}

\date{}

\begin{document}

\begin{titlepage}
\title{
\vskip-40pt
\begin{flushright}
{\small QMUL-PH-10-17}\\
~\\
~\\
~\\
\end{flushright}
{\bf On $\4$ T-duality}
~\\
~\\
\author{Ilya~Bakhmatov\footnote{\tt i.bakhmatov@qmul.ac.uk}
~\\
~\\
~\\
{\it Queen Mary University of London}\\
{\it Centre for Research in String Theory}\\
{\it Department of Physics}\\
{\it Mile End Road, London, E1 4NS, England}
~\\
}}
\maketitle

\begin{abstract}
\noindent
We give a supergravity treatment of the set of bosonic and fermionic T-dualities in the $\4$ background. We consider T-dualities along three flat $AdS_4$ directions, three complexified isometries of $\cp3$, and six fermionic T-dualities. Concentrating on the transformation of the dilaton, we give description of the singularity that arises in the transformation.
\end{abstract}
~\\
~\\

\thispagestyle{empty}
\end{titlepage}

\tableofcontents

\section{Introduction}

Fermionic T-duality is a tree-level symmetry of type II string theory that can be viewed as extending the idea of ordinary T-duality to the superspace setup \cite{bm,brtw}. If one has a Green-Schwarz-type sigma-model that describes the embedding of a string worldsheet in type II superspace, then an analog of the classic Buscher procedure \cite{buscher} can be carried out, resulting in the sigma-model couplings redefinition. The necessary condition is that the background preserves a supersymmetry, parameterized by some Killing spinors $(\e,\ehat)$ (we are considering an $\mathcal{N}=2$ theory, hence a couple of supersymmetry parameters). 

The sigma-model couplings redefinition that results from the fermionic Buscher procedure is quite different from the ordinary T-duality transformation. In fact the entire NS-NS sector is not modified, except for the dilaton that gets an additive contribution
\be\label{f0}
\f' = \f + \frac12 \log C,
\ee
where $C$ is determined by the Killing spinors $(\e,\ehat)$ that parameterize the fermionic isometries, see appendix \ref{app-ferm}. This transformation law is very similar to the way dilaton changes under ordinary T-duality, but the sign of the $\log$ term is opposite. This difference will turn out to be crucial. As for the bosonic fields of the RR sector, their transformation can be written concisely in terms of the bispinor $F^{\a\b}$:
\be
e^{\f'}F'^{\a\b} = e^{\f}F^{\a\b} + k\, \frac{\e^\a\ehat^\b}{C},
\ee
where the precise value of the numerical coefficient $k$ may be different depending on the supergravity conventions, and since we will not be using the RR background transformation in this paper, we do not specify the value of $k$ here. The bispinor $F^{\a\b}$ is formed by contracting all the RR forms of the theory with appropriate antisymmetrized products of gamma-matrices. One can find a more detailed discussion in \cite{bm, ilya, godazgar}.

An important feature of the fermionic T-duality transformation is that it can only be done with complexified Killing spinors, which means that the resulting target space background will generically be a solution to complexified supergravity \cite{ilya}. The paper \cite{godazgar} deals with the extension of fermionic T-duality to a larger class of fermionic symmetries in supergravity, which also include some real transformations.

A crucial ingredient in the proper theoretical understanding of fermionic T-duality would be to formulate it as a group symmetry \cite{fre}, in analogy with the $\mathcal{O}(d,d)$ group representation of the ordinary T-duality. Finally, fermionic T-duality has been recently reformulated as a canonical transformation in phase space \cite{dan}.

In an attempt to further deepen our understanding of the way fermionic T-duality works, in this paper we apply the transformation to an $\4$ background of type IIA string theory. This problem has a rich motivation that comes from the field theory side of the AdS/CFT correspondence. The set of problems related to the Yangian invariance \cite{carlo,plefka} and dual superconformal symmetry \cite{arthur} of scattering amplitudes in ABJM theory \cite{abjm} has attracted much attention recently. Following the $\mathcal{N}=4$ super-Yang-Mills case where the amplitude/Wilson loop correspondence \cite{alday-mald} and the dual superconformal symmetry \cite{drummond,brandhuber} have been proven to exist there are hopes to find and explain similar structures in ABJM theory. In the SYM case the amplitude/Wilson loop correspondence has been explained by a combination of 4+8 T-dualities on the string theory side of the AdS/CFT correspondence. In particular, four ordinary T-dualities along the flat directions of $AdS_5$ and eight fermionic T-dualities were required for the self-duality of the $\5$ background \cite{bm}. Thus, studying the T-duality properties of type IIA string theory in the $\4$ background, which in a certain limit provides the gravity dual to ABJM theory, would in principle do the same to ABJM theory.

However, it has been shown that this approach to dual superconformal symmetry cannot be straightforwardly reproduced in the $\4$ case \cite{adam1,linus,hao}. It is clearly impossible to achieve self-duality in the 3+6 setup (which would be a straightforward mimicking of the $\5$ case) because three ordinary T-dualities would take us from IIA to IIB theory. There has been a proposal \cite{carlo} based on the superalgebra arguments that the correct set of T-dualities to perform in this case would be a `3+3+6' set: three flat $AdS_4$ T-dualities, three $\cp3$ T-dualities, and six fermionic T-dualities. Furthermore, the authors of \cite{arthur} have established the existence of dual superconformal symmetry of the tree-level ABJM scattering amplitudes in case when the dual superspace includes three coordinates corresponding to complexified isometries
of $\mathbb{C}P^3$. Nevertheless, Adam,~Dekel,~and Oz have shown \cite{adam} that this combination of T-dualities is singular. The calculation in \cite{adam} has been done in the supercoset realization of the sigma-model. In the present note we would like to share the complementary point of view on how does this singularity arise. The derivation here is done in terms of the supergravity component fields.

For the sake of simplicity, and also following the conjecture made in \cite{carlo} that the dilaton shifts coming from the bosonic and the fermionic T-dualities seem to cancel, we confine our attention to the transformation of the dilaton. This turns out to be sufficient to expose the nature of the singularity involved. The dilaton gets two additive contributions --- a negative one from the bosonic T-dualitites:
\be
\delta_B\phi = -\frac12 \log |\det g|
\ee
and a positive one from the fermionic dualities:
\be\label{f}
\delta_F\phi = \frac12 \log |\det C|.
\ee
Here $\det g$ is determinant of the block in the metric tensor that incorporates the directions that have been dualized (adapted coordinates have been chosen). An auxiliary function $C$ of (\ref{f0}) is promoted to a matrix because we are considering multiple T-dualities here.

In what follows we shall consider the transformation of the string coupling $e^\f$, which according to the above formulae changes as
\be
\label{result}
e^{2\f'} = e^{2\f} \,\frac{\det C}{\det g}. 
\ee
The main result will be that not only is this transformation singular, but it is also indeterminate, in the sense that {\it both} determinants in the above formula vanish. This is to be contrasted with the $\5$ case \cite{bm}, where the two detereminants are nonzero and cancel precisely, thus allowing for the self-duality.


\section{The background and coordinate systems of $\cp3$}

The $\4$ background has nonzero metric, dilaton, and RR 2- and 4-forms \cite{abjm}:
\bsub
\begin{align}
ds^2 &= \frac{R^3}{k} \left( \frac14 ds^2_{AdS_4} + ds^2_{CP^3} \right),\\
e^{2\phi} &= \frac{R^3}{k^3},\\
\label{F4}F_4 &= \frac{3R^3}{8} \e_4,\\
\label{F2}F_2 &= kJ.
\end{align}
\esub
$ds^2_{AdS_4}$ is a unit radius $AdS_4$ metric, e.g. in the Poincar\'e patch:
\be
\label{ads}
ds^2_{AdS_4} = r^2 \left[-(dx^0)^2+(dx^1)^2+(dx^2)^2 \right]+\frac{dr^2}{r^2},
\ee
and the corresponding 4-form flux $F_4$ is proportional to the totally antisymmetric symbol $\e_4$ in 4 dimensions.

As regards the $\cp3$ part of the background, let us introduce several coordinate systems that will be useful in what follows.
\begin{itemize}
\item Fubini-Study coordinates $(z,\zb)$, where $\zb_\a$ are complex conjugates 
of $z^\a$, $\a = 1,2,3$. Line element has the well-known form
\be
\label{metric-z}
ds^2_{\cp3} = \frac{dz^\a d\zb_\a}{1+|z|^2} - \frac{z^\a \zb_\b 
dz^\b d\zb_\a}{(1+|z|^2)^2},
\ee
where $|z|^2 = z^\a \zb_\a$. The metric is evidently real, which makes it possible to introduce six real coordinates instead.
\item Starting from the real components of the Fubini-Study coordinates $z^\a = \r^\a e^{i\varphi^\a}$, we can introduce six real coordinates $(\m,\a,\th,\ps,\c,\f)$ as follows \cite{hoh}:
\be\label{coords-real}\bn
\r^1 &= \tan\m \,\sin\a \,\sin\frac{\th}{2}, \quad &\varphi^1 &= \frac12\, (\ps - \f + \c), \\
\r^2 &= \tan\m \,\cos\a,                     \quad &\varphi^2 &= \frac12\, \c,              \\
\r^3 &= \tan\m \,\sin\a \,\cos\frac{\th}{2}, \quad &\varphi^3 &= \frac12\, (\ps + \f + \c).
\en\ee
It is convenient to work with the Killing spinors in these coordinates because of the simple representation of the vielbein forms:
\be\bn
e^1 &= d\m,\\
e^2 &= \sin\m \,d\a,\\
e^3 &= \frac12 \sin\m \,\sin\a \left( \cos\ps \,d\th + \sin\th \,\sin\ps \,d\f \right),\\
e^4 &= \frac12 \sin\m \,\sin\a \left( \sin\ps \,d\th - \sin\th \,\cos\ps \,d\f \right),\\
e^5 &= \frac12 \sin\m \,\sin\a \,\cos\a \left( \,d\ps + \cos\th \,d\f \right),\\
e^6 &= \frac12 \sin\m \,\cos\m \left( \,d\c + \sin^2 \a \,d\ps + \sin^2 \a \,\cos\th \,d\f \right).
\en\ee
Line element is simply $ds^2_{\cp3} = \d_{ab} e^a e^b$. We shall use Latin letters for the tangent-space components.
\item Finally, we introduce the complexified $\cp3$ background by means of the following coordinate transformation:
\be\label{coords-w}\bn
w^\a &= z^\a,\\
\wb_\a &= \frac{\zb_\a}{1+|z|^2}.
\en\ee
Since $\wb_\a \neq (w^\a)^*$, we can view them as six independent complex coordinates. The line element takes the simple form:
\be
\label{metric-w}
ds^2_{\cp3} = dw^\a d\wb_\a + \wb_\a \wb_\b dw^\a dw^\b.
\ee
\end{itemize}

The K\"ahler form $J$ in (\ref{F2}) has the simplest representation in the latter coordinates:
\be
\label{kahler}
J = -2i \,dw^\a \wedge d\wb_a.
\ee
Transforming it to the real coordinates, we get 
\be
\label{F2-real}
\bn
J = &-d\m \wedge (d\ps +d\f \cos\th) \sin 2\m \,\sin^2 \a -d\m \wedge d\c \sin 2\m \\
&- d\a \wedge (d\ps + d\f \cos\th) \sin^2 \m \,\sin 2\a + d\th \wedge d\f \sin^2 \m \,\sin^2 \a \,\sin \th.
\en
\ee
This looks much simpler in tangent-space components:
\be
\label{F2-tan}
J_{ab} = e^\m_a e^\n_b J_{\m\n} = 
\left(
	\begin{array}{cccccc}
		& & & & & -2\\
		& & & &-2 &\\
		& & &-2 & &\\
		& &2 & & & \\
		&2 & & & & \\
		2& & & & & 									
	\end{array}
\right).
\ee
\section{Killing vectors}
The six isometries that should be T-dualized are the shifts of three flat $AdS_4$ directions and three internal ($\cp3$) isometries. The contribution of the $AdS_4$ T-dualities can be trivially read off from (\ref{ads}), and it is nonsingular:
\be
\d\f = -3 \log r.
\ee
Therefore from now on we shall only be concerned with internal isometries.

The isometry algebra of $\cp3$ is $\mathfrak{su}(4)$, which is 15-dimensional. None of these isometries commute with any of the supersymmetries, which is the reason for complexifying the Killing vectors. We use the complexified Killing vectors of $\cp3$ as given in \cite{pope}:
\be\label{pope}
\bn
K^\a &= {T_0}^\a + {T_\b}^\a z^\b - {T_0}^0 z^\a - {T_\b}^0 z^\b z^\a,\\
K_\a &= -{T_\a}^0 - {T_\a}^\b \zb_\b + {T_0}^0 \zb_\a + {T_0}^\b \zb_\b \zb_\a,
\en
\ee
for a vector field
\be
K = K^\a \,\frac{\p}{\p z^\a} + K_\a \,\frac{\p}{\p \zb_\a}.
\ee
There are precisely 15 independent parameters ${T_A}^B$, $A,B = 0,\ldots,3$ because they are subject to the constraint ${T_A}^A = 0$.

We shall consider the three complex Killing vectors that result from keeping ${T_0}^\a$ in (\ref{pope}):
\be
\label{KV}
K_{(\a)} = \frac{\p}{\p z^\a} + \zb_\a \zb_\b \frac{\p}{\p \zb_\b},\quad \a=1,2,3.
\ee
These three vectors commute with each other and by transforming them to the real coordinates (\ref{coords-real}) one can check that they are of the form $a+ib$, where $a$ and $b$ are ordinary real Killing vectors of $\cp3$:
\bsub
\label{KV-real}
\be\bn
K_{(1)} = &\frac12 \,e^{-\frac{i}{2} (\ps-\f+\c)} \left( \sin\a \sin\frac{\th}{2} \frac{\p}{\p\m} + \cot\m \cos\a \sin\frac{\th}{2} \frac{\p}{\p\a} + \frac{\cot\m \cos\frac{\th}{2}}{\sin\a} \frac{\p}{\p\th}\right.
\\
&\left. -i \,\frac{\cot\m}{\sin\a \sin\frac{\th}{2}} \frac{\p}{\p\ps} +i \,\frac{\cot\m}{\sin\a \sin\frac{\th}{2}} \frac{\p}{\p\f} +2i \,\tan\m \sin\a \sin\frac{\th}{2} \frac{\p}{\p\c}\right),
\en
\ee
\be\bn
K_{(2)} = \frac12 \,e^{-\frac{i}{2} \c} &\left[ \cos\a \,\frac{\p}{\p\m} - \cot\m \sin\a \,\frac{\p}{\p\a} +2i \,\frac{\cot\m}{\cos\a} \,\frac{\p}{\p\ps}\right. 
\\
&\left. -2i \left( \frac{\cot\m}{\cos\a} - \frac{\cos\a}{\cot\m} \right) \frac{\p}{\p\c}\right],
\en
\ee
\be\bn
K_{(3)} = &\frac12 \,e^{-\frac{i}{2} (\ps+\f+\c)} \left( \sin\a \cos\frac{\th}{2} \,\frac{\p}{\p\m} + \cot\m \cos\a \cos\frac{\th}{2} \,\frac{\p}{\p\a} -2 \,\frac{\cot\m \sin\frac{\th}{2}}{\sin\a} \,\frac{\p}{\p\th}\right.
\\
&\left. -i \,\frac{\cot\m}{\sin\a \cos\frac{\th}{2}} \,\frac{\p}{\p\ps} -i \,\frac{\cot\m}{\sin\a \cos\frac{\th}{2}} \,\frac{\p}{\p\f} +2i \,\tan\m \sin\a \cos\frac{\th}{2} \,\frac{\p}{\p\c}\right).
\en
\ee
\esub

Note that alternatively one could also use the three vector fields corresponding to ${T_\a}^0$, which are complex conjugates of the vectors (\ref{KV}), or those resulting from keeping ${T_\a}^\a$ (no sum). These two groups of complex Killing vectors also commute among themselves.

Now we can reveal the reason for the introduction of the $(w,\wb)$ coordinates in (\ref{coords-w}). Transforming the vectors (\ref{KV}) to these coordinates one discovers that they are acting as shifts \footnote{This has been pointed out to the author by Carlo Meneghelli.}:
\be
K_{(\a)} = \frac{\p}{\p w^\a},
\ee
which enables us to calculate $\det g$ in (\ref{result}). For this purpose, we read off the metric tensor from the expression for the interval in $(w,\wb)$ coordinates (\ref{metric-w}):
\be
\label{metric}
g_{\m\n} = \left(
	\begin{array}{cc}
		\begin{array}{ccc}
			\wb^1 \wb^1 & \wb^1 \wb^2 & \wb^1 \wb^3\\
			\wb^2 \wb^1 & \wb^2 \wb^2 & \wb^2 \wb^3\\
			\wb^3 \wb^1 & \wb^3 \wb^2 & \wb^3 \wb^3\\						
		\end{array}	& 
		\begin{array}{ccc}
			  &   &  \\
			  & 1/2 &  \\
			  &   &  \\						
		\end{array} \\
		\begin{array}{ccc}
			  &   &  \\
			  & 1/2 &  \\
			  &   &  \\						
		\end{array} &
		\begin{array}{ccc}
			  &   &  \\
			  & 0 &  \\
			  &   &  \\						
		\end{array} \\
	\end{array}\right).
\ee
The upper-left block here corresponds to the $dw\,dw$ term in the interval. Rescaling of the string coupling under the three T-dualities with respect to $K_{(1,2,3)}$ is given by the determinant of this block, which is identically zero. Now we can rewrite (\ref{result}) as
\be
e^{2\f'} = e^{2\f} \,\frac{\det C}{0}. 
\ee
This is clearly a singularity, and now we proceed to showing that the numerator in this formula vanishes as well.

\section{Killing spinors}

In order to get an expression for the matrix $C$ (\ref{2a-c}) we need to know the Killing spinors $\e,\ehat$. These can be found as solutions to the equations
\bsub
\begin{align}
\label{Ksp1}&\left( \slashed{F}_2  - \frac13 \slashed{F}_4 \G^{11} \right) \mathrm{E} = 0,\\
\label{Ksp2}&\nabla_M \mathrm{E} = \frac{e^\phi}{8} \left( \slashed{F}_2 \G_M \G^{11} - \slashed{F}_4 \G_M \right) \mathrm{E},
\end{align}
\esub
which are conditions that supersymmetry variations of the type IIA fermions vanish. Supersymmetry parameter $\mathrm{E}$ is a Majorana spinor, while $\e$ and $\ehat$ are its Majorana-Weyl components, which can be obtained by applying the projections $\frac12 (1\pm \G^{11})$. We use the notation $\slashed{F}_n = \frac{1}{n!} F_{M_1\ldots M_n} \G^{M_1\ldots M_n}$. Note that the free index $M$ in (\ref{Ksp2}) is a curved index.

Original derivation of the Killing spinors of $\cp3$ can be found in \cite{nilpope}, \cite{hoh}, and \cite{hikida}. Here we shall briefly overview the derivation for the sake of consistency with our notation and conventions. We decompose the spinor parameter $\mathrm{E} = \k \otimes \h$ into the product of the $SO(1,3)$ and $SO(6)$ spinors $\k$ and $\h$. With the corresponding decomposition of the gamma-matrices (for details see appendix \ref{app-gamma}), the first Killing spinor equation (\ref{Ksp1}) becomes
\be
\left( 1\otimes \frac12 F_{ij} \b^{ij} \right)\,\left(\k\otimes\h\right) = \left( 1\otimes 2 \b^7 \right)\,\left(\k\otimes\h\right).
\ee
We see that $\k$ is unconstrained, while the equation for $\h$ can be rewritten as follows:
\be
\label{eigen}
Q \,\b^7 \h = -2 \,\b^7 \h,
\ee
where $Q = \frac12 F_{ij} \b^{ij} \b^7$. Evaluating this matrix operator using the tangent-space components of the 2-form (\ref{F2-tan}) shows that indeed there is a~$-2$ eigenvalue, whose degeneracy is~$6$. The corresponding 6-parameter eigenspinor has the form
\be\label{eta}
\h = \left(\begin{array}{cccccccc}-f_1 & f_2 & f_3 & -f_1 & f_4 & -f_5 & -f_6 & f_4\end{array}\right)^T.
\ee
The exact functional dependence of the parameters $f_i$ on spacetime coordinates is fixed by the second Killing spinor equation (\ref{Ksp2}).

Performing the same decomposition as above we arrive at the following equations for $\k$ and $\h$:
\be
\left(\p_{\underline\m} + \frac14 \,\w_{\underline\m,\r\l} \,\a^{\r\l}\right)\k = \a_{\underline\m} \,\a^5\k,\\
\ee
\be
\left(\p_{\underline i} + \frac14 \,\w_{\underline i,kl} \,\b^{kl}\right)\h = \frac{i}{2} \b_{\underline i} \,\h - \frac{i}{4} F_{\underline i\,j} \,\b^j \b^7\,\h,
\ee
where we underline the world indices and our convention for the spin connection is 
\be
\w_{\underline A,BC} = \frac12 \,e^{\underline D}_B \,e^{\underline E}_C \left( \W_{\underline {ADE}} -\W_{\underline {DEA}}  +\W_{\underline {EAD}} \right),
\ee
\be
\W_{\underline{ABC}} = \p_{\left[\underline{A}\right.} e^D_{\left.\underline{B}\right]} \,e^E_{\underline{C}} \,\h_{DE}.
\ee

The $AdS_4$ Killing spinor equation is easy to solve and the solution $\k$ is 4-parametric:
\be\label{kappa}
\k =\left(
\begin{array}{c}
\k_1 r^{-1/2} \\ \k_2 r^{-1/2} \\ r^{1/2} \left[ -\k_2 (x^0-x^1) + \k_1 x^2 +\k_3 \right] \\ r^{1/2} \left[ \k_1 (x^0+x^1) - \k_2 x^2 +\k_4 \right]
\end{array}
\right).
\ee
Solving the equations for $\h$ is more tedious, but it can be done analytically. The solution is very bulky and is therefore given in the appendix \ref{app-spinors}. The overall result is that the $AdS_4$ part of the Killing spinor $\k$ is 4-parametric, while the $\cp3$ part is 6-parametric. Thus there are $24$ independent Killing spinors in the $\4$ background.

\section{Symmetry superalgebra}

We now need to establish which Killing spinors to use for the T-duality transformation. As long as we have chosen the three isometries generated by (\ref{KV}), the choice of the fermionic symmetries is dictated by the requirement that together they form a commuting subalgebra of the symmetry superalgebra. Bosonic generators (\ref{KV}) of this subalgebra are commuting; our next step will be to select the fermionic generators (Killing spinors) that commute with these three vectors and finally we shall check the anticommutation of the selected supersymmetries among themselves.

First of all recall that apart from (\ref{KV}) our T-duality setup includes three bosonic dualities along the flat directions of $AdS_4$. Looking at the $AdS_4$ part of the Killing spinor (\ref{kappa}) we see that we must set $\k_{1,2} = 0$ for the product $\k\otimes\h$ to be invariant under the shifts of $x^{0,1,2}$. So what happens to the $\cp3$ part of the Killing spinor?

From the explicit expressions of the $\cp3$ spinors (appendix \ref{app-spinors}) it is not easy to tell what are their commutation properties with the vectors (\ref{KV}). Therefore we calculate the Lie derivatives of our Killing spinor fields with respect to the Killing vectors \cite{fig}. Lie derivative of a spinor $\h$ with respect to a vector $K$ is given by
\be
\mathcal{L}_K \h = K^i \nabla_i \h + \frac12 \nabla_{\left[i\right.} K_{\left.j\right]} \,\frac12 \b^{ij} \h,
\ee
where of course the covariant derivatives of a vector and of a spinor are taken correspondingly with respect to the Christoffel and spin connections. 

Using the expressions for $K_{(1,2,3)}$ (\ref{KV-real}) and for $\h_{1,\ldots,6}$ ((\ref{eta}) and appendix \ref{app-spinors}, where the spinor $\h_i$ results from keeping only the parameter $h_i = 1$ and setting all the rest to zero), one finds the following algebra:
\bsub
\be\bn
\mathcal{L}_{K_{(1)}} \h_1 &= -\frac{1}{2} (\h_3 - i\h_4 + i\h_5 - \h_6),\\
\mathcal{L}_{K_{(1)}} \h_2 &= -\frac{i}{2} (\h_3 - i\h_4 + i\h_5 - \h_6),\\
\mathcal{L}_{K_{(1)}} \h_3 &=  \frac{1}{4} (\h_1 + i\h_2               ),\\
\mathcal{L}_{K_{(1)}} \h_4 &= -\frac{i}{4} (\h_1 + i\h_2               ),\\
\mathcal{L}_{K_{(1)}} \h_5 &=  \frac{i}{4} (\h_1 + i\h_2               ),\\
\mathcal{L}_{K_{(1)}} \h_6 &= -\frac{1}{4} (\h_1 + i\h_2               ),\\
\en\ee
\be\bn
\mathcal{L}_{K_{(2)}} \h_1 &= 0,\\
\mathcal{L}_{K_{(2)}} \h_2 &= 0,\\
\mathcal{L}_{K_{(2)}} \h_3 &= -\frac{i}{2} (\h_4 - i\h_6               ),\\
\mathcal{L}_{K_{(2)}} \h_4 &=  \frac{i}{2} (\h_3 + i\h_5               ),\\
\mathcal{L}_{K_{(2)}} \h_5 &=  \frac{1}{2} (\h_4 - i\h_6               ),\\
\mathcal{L}_{K_{(2)}} \h_6 &=  \frac{1}{2} (\h_3 + i\h_5               ),\\
\en\ee
\be\bn
\mathcal{L}_{K_{(3)}} \h_1 &= -\frac{i}{2} (\h_3 + i\h_4 + i\h_5 + \h_6),\\
\mathcal{L}_{K_{(3)}} \h_2 &=  \frac{1}{2} (\h_3 + i\h_4 + i\h_5 + \h_6),\\
\mathcal{L}_{K_{(3)}} \h_3 &=  \frac{i}{4} (\h_1 + i\h_2               ),\\
\mathcal{L}_{K_{(3)}} \h_4 &= -\frac{1}{4} (\h_1 + i\h_2               ),\\
\mathcal{L}_{K_{(3)}} \h_5 &= -\frac{1}{4} (\h_1 + i\h_2               ),\\
\mathcal{L}_{K_{(3)}} \h_6 &=  \frac{i}{4} (\h_1 + i\h_2               ).\\
\en\ee
\esub
It is easy to see that there are three linear combinations of the Killing spinors that are invariant under the action of all three vectors:
\be
\h_1 + i\h_2,\quad \h_3 + i\h_5,\quad \h_4 - i\h_6.
\ee
Tensor multiplying these with the two $AdS_4$ spinors ($\k_3, \k_4 \neq 0$ in (\ref{kappa})) we get the six Killing spinors, which is precisely the number needed for the T-duality. Thus the symmetry superalgebra constraints unambiguously fix the fermionic directions to be T-dualized.

It remains to make sure that the corresponding supersymmetries anticommute. The constraint on the spinor $\mathrm{E} = \k\otimes\h$ is given in the appendix (\ref{2a-constraint}) and can be checked straightforwardly. For multiple supersymmetries one has to generalize this to the matrix constraint
\be
\bar{\mathrm{E}}_i \,\G^\m \mathrm{E}_j = 0, \quad i,j = 1,\ldots,6.
\ee

\section{Fermionic T-duality}

Finally we are in a position to calculate the matrix $C_{ij}$, $i,j=1,\ldots,6$:
\be
\p_\m C_{ij} = \bar{\mathrm{E}}_i \,\G_\m \G^{11}\mathrm{E}_j,
\ee
which is a generalisation of (\ref{2a-c}) for the case of multiple T-dualities. These equations turn out to be consistent, and the solution is (up to integration constants)
\be\label{C2}
C_{\4} = \left(
			\begin{array}{cc}
				0 & \begin{array}{ccc}
							0 & a & b\\
						   -a & 0 & c\\
						   -b & -c & 0
					   \end{array}\\
				\begin{array}{ccc}
							0 & -a & -b\\
						    a & 0 & -c\\
						    b & c & 0
					   \end{array} & 0
			\end{array}
		\right) 
,
\ee
where
\bsub
\label{abc}
\begin{align}
a &= -2\, r\, e^{-\frac{i}{2} (\ps +\c)} \sin{2\m} \sin{\a} \left[ \cos{\frac12 (\th +\f)} + i \sin{\frac12 (\th -\f)} \right],\\
b &=  2\, r\, e^{-\frac{i}{2} (\ps +\c)} \sin{2\m} \sin{\a} \left[ i \cos{\frac12 (\th -\f)} + \sin{\frac12 (\th +\f)} \right],\\
c &= -2\, r\, e^{-\frac{i}{2} \c} \sin{2\m} \cos{\a}.
\end{align}
\esub

The important point to notice here is that the determinant of the matrix (\ref{C2}) is identically zero, irrespective of the values (\ref{abc}). This is the second singularity, which manifests itself in the numerator of the formula (\ref{result}). 

The vanishing of $\det C$ in the present case is to be contrasted with the $\5$ case \cite{bm}, where the $C$-matrix has the same algebraic structure (symmetric matrix with off-diagonal antisymmetric blocks). However since in this setup one does $4$ bosonic ($AdS$) dualities and $8$ fermionic ones, $C_{\5}$ is now an $8 \times 8$ matrix:
\be\label{C3}
C_{\5} = \left(
			\begin{array}{cc}
				0 & \begin{array}{cccc}
							0 & a & b & c\\
						   -a & 0 & d & e\\
						   -b & -d & 0 & f\\
						   -c & -e & -f & 0
					   \end{array}\\
				\begin{array}{cccc}
							0 & -a & -b & -c\\
						    a & 0 & -d & -e\\
						    b & d & 0 & -f\\
							c & e & f & 0
					   \end{array} & 0
			\end{array}
		\right)
,
\ee
where the entries are given by 
\bsub
\begin{align}
a &= 2 R r \sin y^1 \left(\cos y^2 - i \sin y^2 \cos y^3\right),\\
b &= 2 R r \left(i \cos y^1 + \sin y^1 \ldots \sin y^{5}\right),\\
c &= -2 R r \sin y^1 \sin y^2 \sin y^3 \left( \cos y^4 - i \sin y^4 \cos y^{5}\right),\\
d &=  2 R r \sin y^1 \sin y^2 \sin y^3 \left( \cos y^4 + i \sin y^4 \cos y^{5}\right),\\
e &= 2 R r \left(-i \cos y^1 + \sin y^1 \ldots \sin y^{5}\right),\\
f &= -2 R r \sin y^1 \left(\cos y^2 + i \sin y^2 \cos y^3\right).
\end{align}
\esub
Here $r$, as before, is the $AdS$ radial coordinate, $R$ is the $AdS$ radius and the variables $\{y^1,\ldots,y^5\}$ are the standard coordinates on $\mathbb{S}^5$:
\be
ds^2 = (dy^1)^2 + \sin^2 y^1 \left\{ (dy^2)^2 + \sin^2 y^2 \left[ (dy^3)^2 + \ldots\right]\right\}.
\ee
In the $\5$ case, not only is $\det C_{\5}$ nonvanishing, but for these particular values of the entries it can be simplified to
\be
\det C_{\5} = (2 R r)^8.
\ee
This is precisely cancelled by the $4$ $AdS$ dualities.

\section{Summary}

We have shown that under the combination of bosonic and fermionic T-dualities in the directions given by the three complexified $\cp3$ isometries and six complexified supersymmetries the transformation of the dilaton is indeterminate:
\be
e^{2\f'} = e^{2\f} \,\frac{0}{0}. 
\ee
This provides an alternative point of view on T-dualizing $\4$ background that has been done recently by Adam, Dekel, and Oz~\cite{adam} in the supercoset formulation of the sigma-model. Perhaps a way to eliminate this ambiguity would be to consider a deformed $\4$ background, the deformation being parameterized by some $\l$, such that the dependence on the deformation parameter $e^{2\f'} = f(\l)\,e^{2\f}$ would have a well-defined limit when the deformation is removed $\lim_{\l\rightarrow 0} f(\l)$.

Most likely this deformation would require giving the dilaton some nontrivial coordinate dependence. The dilaton equation of motion in our conventions is 
\be
R = 4(\p\f)^2 - 4\nabla^2 \f
\ee
(for a vanishing $B$-field). If we keep the dilaton constant, then the requirement that the $AdS$ part of the geometry be preserved will only allow for the deformations of the $\mathbb{C}P^3$ part that preserve $R=0$, which is problematic. One can also consider the Killing spinor equation (\ref{Ksp1}), which in the ABJM background reduces to the eigenspinor condition (\ref{eigen}). If one were to deform the RR 2-form, the eigenspinor condition would be broken, and for some supersymmetry to be preserved one would have to introduce the dilaton into the game. With nontrivial dilaton the equation (\ref{eigen}) gets modified to
\be
\left[ k \b^i \p_i \f - e^\f \left( Q + 2 \right) \right] \b^7 \h= 0,
\ee
where, as before, $Q = \frac12 F_{ij} \b^{ij} \b^7$, and we have absorbed the numerical factors that depend on the supergravity conventions into the constant $k$. An appropriate relative normalization of $F_2$ and $F_4$ is also assumed. It is possible that the dilaton field with nontrivial dependence on the internal manifold could allow for some supersymmetry to be preserved under the deformation.

A candidate recipe for the deformation is the TsT transformation \cite{lunin,laura}, which gives the beta-deformed $\4$ theory described in \cite{imeroni}. In order for the Killing vectors to be preserved under the beta-deformation, one may carry out the beta-deformation with respect to these Killing vectors. Therefore if we beta-deform the $\4$ background using the directions (\ref{KV}), we can then use the same Killing vectors for the T-duality. However the $dw\,dw$ block in (\ref{metric}) is not affected by such a beta-deformation, which means that the corresponding determinant is still zero. Thus the use of the TsT transformation for the deformation purposes in our setup is problematic.

A general property of fermionic T-duality that has been revealed in the present work is that the transformation may be singular. In the case at hand the degeneracy of the matrix (\ref{C2}) as opposed to (\ref{C3}) is due to their block structure with antisymmetric blocks (an odd-dimensional antisymmetric matrix has zero determinant). In a different setup the structure of the $C$-matrix may be different \cite{ilya} (as one can observe from the definition (\ref{2a-c}), $C$ is only required to be symmetric). It is yet to be understood what makes singular transformations possible, and in particular what is the role of complexification of the fermionic symmetries that is obligatory for doing fermionic T-duality.

\section*{Acknowledgements}

The author would like to thank Nathan Berkovits, David Berman, Stefan Hohenegger, and Carlo Meneghelli for valuable discussions.

This work is supported by Westfield Trust Scholarship.

\appendix

\section{Fermionic T-duality in type IIA}
\label{app-ferm}

Start with any Majorana-Weyl representation of the gamma-matrices, such that
\be
\G^\m =
	\left(
        \begin{array}{cc}
            0 & (\g^\m)^{\a\b}\\
            \g^\m_{\a\b} & 0\\
        \end{array}
    \right), \quad
C = 
	\left(
        \begin{array}{cc}
            0 & {c_\a}^\b\\
            \bar{c}^\a{}_\b & 0\\
        \end{array}
    \right), \quad
\G^{11} =
	\left(
        \begin{array}{cc}
            1 & 0\\
            0 & -1\\
        \end{array}
    \right).
\ee
The indices $\a, \b$ here take values $1\ldots 16$. Different properties of this class of representations are considered in \cite{van}. We use Majorana conjugation for covariant spinors $\bar\ps = \ps^T C$.

In type IIA it is possible to write the main formulae of fermionic T-duality \cite{bm} concisely by introducing a single Majorana spinor parameter instead of a pair $(\e,\ehat)$ of Majorana-Weyl spinors. In particular, the abelian constraint for a supersymmetry generated by $(\e,\ehat)$ is
\be\bn
0 =\left( \bar\e \,Q + \bar\ehat \,\hat{Q} \right)^2 &= - \left( \bar\e \,\G^\m \e + \bar\ehat \,\G^\m \ehat \right) P_\m 
\\
  &= \left\{ 
		\begin{array}{l}
			\mathrm{IIB}:\quad \left[(\e c)^\a \g^\m_{\a\b} \e^\b + (\ehat c)^\a \g^\m_{\a\b} \ehat^\b\right] P_\m, \\
			\mathrm{IIA}:\quad \left[(\e c)^\a \g^\m_{\a\b} \e^\b + (\ehat \bar{c})_\a (\g^\m)^{\a\b} \ehat_\b\right] P_\m.
		\end{array}
	\right.
\en\ee
The IIA expression can be rewritten in terms of the Majorana spinor $\mathrm{E} = \e + \ehat$:
\be
\label{2a-constraint}
\bar{\mathrm{E}} \,\G^\m \mathrm{E} = 0.
\ee

Another relation of interest is the definition of an auxiliary function $C$ (not to be confused with the charge conjugation matrix):
\be
\p^\m C = i \left( \bar\e \,\G^\m \e - \bar\ehat \,\G^\m \ehat \right) = \left\{ 
\begin{array}{l}
	\mathrm{IIB}:\quad i\left[(\e c)^\a \g^\m_{\a\b} \e^\b - (\ehat c)^\a \g^\m_{\a\b} \ehat^\b\right], \\
	\mathrm{IIA}:\quad i\left[(\e c)^\a \g^\m_{\a\b} \e^\b - (\ehat \bar{c})_\a (\g^\m)^{\a\b} \ehat_\b\right].
\end{array}
\right.
\ee
Again we can rewrite the IIA expression succinctly as
\be
\label{2a-c}
\p^\m C = \bar{\mathrm{E}} \,\G^\m \G^{11}\mathrm{E}.
\ee

Since the expressions in (\ref{2a-constraint}) and (\ref{2a-c}) are vectors, these formulae are independent of the gamma-matrix representation.

\section{Gamma-matrices}
\label{app-gamma}
For the purposes of working with type IIA supergravity, whose spinorial quantities are Majorana spinors of $1+9$-dimensional spacetime, we need a Majorana representation of the gamma-matrices. This will be built as a product of Majorana representations in $1+3$ and in $6$ dimensions.

Our spacetime signature convention is $(- + \ldots +)$, hence the following four real anticommuting matrices $\a^\m$ furnish a Majorana representation in $D=1+3$:
\be
\begin{array}{ccccc}
\a^0 & = & \s_3 & \otimes & i\s_2,\\
\a^1 & = & \s_3 & \otimes & \s_1,\\
\a^2 & = & \s_3 & \otimes & \s_3,\\
\a^3 & = & \s_1 & \otimes & 1.
\end{array}
\ee
Volume element $\a^5 = \a^0 \ldots \a^3 = i\s_2 \otimes 1$ is also real and squares to $-1$.

We choose the six gamma-matrices of $6D$ Euclidean space to be
\be
\begin{array}{ccccccc}
\b^1 & = & 1    & \otimes & \s_2 & \otimes & \s_1,\\
\b^2 & = & 1    & \otimes & \s_2 & \otimes & \s_3,\\
\b^3 & = & \s_1 & \otimes & 1    & \otimes & \s_2,\\
\b^4 & = & \s_3 & \otimes & 1    & \otimes & \s_2,\\
\b^5 & = & \s_2 & \otimes & \s_1 & \otimes & 1,   \\
\b^6 & = & \s_2 & \otimes & \s_3 & \otimes & 1.
\end{array}
\ee
These are imaginary and we define the corresponding volume element to be real: $\b^7 = -\b^1\ldots\b^6 = i\s_2\otimes i\s_2\otimes i\s_2$.

Finally, the ten-dimensional real gamma-matrices $\G$ are the following products (for $\a = 0,\ldots,3$ and $i = 1,\ldots,6$):
\be
\begin{array}{ccccc}
\G^\m & = & \a^\m & \otimes & 1,\\
\G^{i+3}  & = & i\a^5 & \otimes & \b^i
\end{array}
\ee
Ten-dimensional chirality operator is $\G^{11} = \G^0 \ldots \G^9 = -\a^5 \otimes \b^7$. This representation is clearly not Weyl.

\section{$\cp3$ Killing spinors}
\label{app-spinors}

The components of the $\cp3$ factor (\ref{eta}) of the Killing spinor are given by the following:
\be\bn
f_1 = \frac{1}{2} &\left\{2 h_1 \cos\a \sin\frac{\c}{4} + 
	2 h_2 \cos\a \cos\frac{\c}{4} +\right.
	\\   
   &\sin\a \left[\left(h_3 \sin\frac{\f}{2}+h_4
   \cos\frac{\f}{2}\right) \sin\frac{1}{4} (2 \th +\c -2 \ps)+\left(h_3 \cos\frac{\f}{2}
   -h_4 \sin\frac{\f}{2}\right) \right.
   \\
   &\cos\frac{1}{4} (2 \th -\c +2 \ps)-\left(h_6
   \cos\frac{\f}{2}-h_5 \sin\frac{\f}{2}\right) \cos
   \frac{1}{4} (2 \th +\c -2 \ps)
   \\
   &+\left.\left.\left(h_6 \sin\frac{\f}{2}+h_5 \cos\frac{\f}{2}\right) 
   \sin\frac{1}{4} (2 \th -\c +2 \ps)\right]\right\},
\en\ee
\be\bn
f_4 = \frac{1}{2} &\left\{2 h_1 \cos\a \cos\frac{\c}{4} - 2 h_2 \cos\a \sin\frac{\c}{4} +\right.
	\\   
   &\sin\a \left[\left(h_3 \sin\frac{\f}{2}+h_4
   \cos\frac{\f}{2}\right) \cos\frac{1}{4} (2 \th +\c -2 \ps)+\left(h_3 \cos\frac{\f}{2}
   -h_4 \sin\frac{\f}{2}\right) \right.
   \\
   &\sin\frac{1}{4} (2 \th -\c +2 \ps)+\left(h_6
   \cos\frac{\f}{2}-h_5 \sin\frac{\f}{2}\right) \sin
   \frac{1}{4} (2 \th +\c -2 \ps)
   \\
   &-\left.\left.\left(h_6 \sin\frac{\f}{2}+h_5 \cos\frac{\f}{2}\right) 
   \cos\frac{1}{4} (2 \th -\c +2 \ps)\right]\right\},
\en\ee
\be\bn
f_2 = \frac{1}{2} &\Big\{\mathcal{A}_1\big[(\cos\a +1) \sin\m-(\cos\a -1) \cos\m\big] + \mathcal{B}_1\big[(\cos\a +1) \cos\m -(\cos\a -1) \sin\m\big]
   \\
   &-2
   \sin\a (\cos\m -\sin\m) \left(h_1 \cos\frac{\c}{4}-h_2
   \sin\frac{\c}{4}\right)\Big\},
\en\ee
\be\bn
f_3 = \frac{1}{2} &\Big\{\mathcal{A}_1\big[(\cos\a -1) \sin\m+(\cos\a +1) \cos\m\big] + \mathcal{B}_1\big[(\cos\a -1) \cos\m +(\cos\a +1) \sin\m\big]
   \\
   &+2
   \sin\a (\cos\m +\sin\m) \left(h_1 \cos\frac{\c}{4}-h_2
   \sin\frac{\c}{4}\right)\Big\},
\en\ee
\be\bn
f_5 = \frac{1}{2} &\Big\{\mathcal{A}_2\big[(\cos\a +1) \sin\m-(\cos\a -1) \cos\m\big]
	+
   \mathcal{B}_2\big[(\cos\a +1) \cos\m -(\cos\a -1) \sin\m\big]
   \\
   &-2
   \sin\a (\cos\m -\sin\m) \left(h_1 \cos\frac{\c}{4}-h_2
   \sin\frac{\c}{4}\right)\Big\},
\en\ee
\be\bn
f_6 = \frac{1}{2} &\Big\{\mathcal{A}_2\big[(\cos\a -1) \sin\m+(\cos\a +1) \cos\m\big] + \mathcal{B}_2\big[(\cos\a -1) \cos\m +(\cos\a +1) \sin\m\big]
   \\
   &+2
   \sin\a (\cos\m +\sin\m) \left(h_1 \cos\frac{\c}{4}-h_2
   \sin\frac{\c}{4}\right)\Big\},
\en\ee
where
\be\bn
\mathcal{A}_1 &=\left[\sin
   \frac{\ps}{2} \sin\frac{1}{4} (2 \th -\c) \left(h_3 \cos
   \frac{\f}{2}-h_4 \sin\frac{\f}{2}\right)-\sin
   \frac{\ps}{2} \cos\frac{1}{4} (2 \th -\c) \right.
\\   
   &\left(h_6 \sin
   \frac{\f}{2}+h_5 \cos\frac{\f}{2}\right)+\cos
   \frac{\ps}{2} \left(\cos\frac{1}{4} (2 \th +\c)
   \left(h_6 \cos\frac{\f}{2}-h_5 \sin\frac{\f}{2}\right)\right.
   \\
   &-\left.\left.\sin\frac{1}{4} (2 \th +\c) \left(h_3 \sin
   \frac{\f}{2}+h_4 \cos\frac{\f}{2}\right)\right)\right],\\
\mathcal{B}_1 &=  \left(\cos\frac{\ps}{2} \cos\frac{1}{4} (2 \th -\c
   ) \left(h_3 \cos\frac{\f}{2}-h_4 \sin\frac{\f
   }{2}\right)+\cos\frac{\ps}{2} \sin\frac{1}{4} (2 \th -\c) \right.
	\\   
   &\left(h_6 \sin\frac{\f}{2}+h_5 \cos\frac{\f
   }{2}\right)-\sin\frac{\ps}{2} \left(\cos\frac{1}{4} (2 \th
   +\c) \left(h_3 \sin\frac{\f}{2}+h_4 \cos\frac{\f
   }{2}\right)\right.
\\   
   &+\left.\left.\sin\frac{1}{4} (2 \th +\c) \left(h_6 \cos
   \frac{\f}{2}-h_5 \sin\frac{\f}{2}\right)\right)\right),
\en\ee
and $\mathcal{A}_2, \mathcal{B}_2$ are the same with the following substitution: 
\be\bn
\sin\frac{\ps}{2} &\rightarrow -\cos\frac{\ps}{2},\\
\cos\frac{\ps}{2} &\rightarrow \sin\frac{\ps}{2}.
\en\ee

\bibliographystyle{utphys}
\bibliography{1}

\end{document}